\documentstyle[PASJadd,psfig]{PASJ95}
\draft
\begin{document}

\title{ASCA Observation of the Lyman-limit Quasar PKS2145+067}
\author{Noriko Y. {\sc Yamasaki}, Teru {\sc Ishikawa} and Takaya {\sc Ohashi} \\
          {\it Department of Physics, Tokyo Metropolitan University, 
           1-1 Minami-Ohsawa, Hachioji, Tokyo 192-0397} \\
           {\it E-mail(NYY):yamasaki@phys.metro-u.ac.jp} } 
\abst{
X-ray  observation of a famous Lyman-limit quasar  PKS2145+067
at $z_{\rm em}=$ 0.990  
was carried with {\it ASCA}.
The source showed  a 2-10 keV flux of 1.3$\times$10$^{-11}$ 
erg cm$^{-2}$ sec$^{-1}$ 
($L_{\rm X} = 2.5 \times 10^{46}$ erg s$^{-1}$ 
for $H_{0}=$ 50 km s$^{-1}$ Mpc$^{-1}$) 
described by a power-law spectrum with a photon index 
 $\Gamma$=1.63$\pm0.04$ .
In the {\it ASCA} energy band, no excess absorption 
was  detected implying  the absorption column density 
at $z_{\rm ab}=$0.791 was  less than 1.6$\times$10$^{21}$ cm$^{-2}$ 
if absorbing medium had  a metal abundance of 0.5 solar.
Comparison with previous {\it Einstein} and {\it ROSAT} 
 observations shows that  PKS2145+067 has increased  
its luminosity by a  factor of 2$\sim$3 between 1991 and 1998.
}
\kword{Galaxies:intergalactic medium --- Quasars--- X-rays:spectra}
\maketitle
\thispagestyle{headings}
\section{Introduction}
PKS2145+067 (4C +06.69, 
$[\alpha_{2000},\delta_{2000}]=[$21h48m05s,06$^{\circ}$57'39'']) 
is an X-ray luminous AGN at $z_{\rm em}$=0.990
and is known to show a Lyman limit system at $z_{\rm abs}=$0.791 
which accompanies  many metal lines indicating low and 
high ionization states 
(C {\small III}, C {\small IV}, N {\small III}, N {\small V}, O {\small VI}, 
Si {\small III}, Si {\small IV}, and Mg {\small II}) 
(Bahcall et al. 1993, Bergeron et al. 1994).
The source  is a radio-loud QSO and is 
classified as a flat-spectrum radio quasars (FSRQ) 
(see ex. Padovani 1992).
The polarization of this quasar is as low as 1.0$\pm$0.6 \% 
(Visvanathan \& Wills 1998), and a submilliarcsecond imaging 
by the Very Long Baseline Array (VLBA) at 15GHz does not detect any 
jet-like structure (Kellermann et al. 1998). 

The  Lyman limit system exhibits  a  neutral hydrogen column density 
 as large as 
$N$(HI)=3.2$\times$10$^{17}$ cm$^{-2}$ (1.6$\times$10$^{16} \leq N$(HI)$\leq$
6.3$\times$10$^{17}$ cm$^{-2}$) with a Doppler parameter $b$ of 
$35\leq b \leq 63$ km s$^{-1}$ (Bergeron et al. 1994).
An associated galaxy is identified at 55$h_{0}^{-1}$ kpc away 
from the line of sight (Bergeron \& Boiss\`{e} 1991) and is 
considered to be in star formation activity. 
The co-existence of the absorption lines from 
a low-ionized element Mg {\small II} and a high-ionized O {\small VI} 
suggest that 
there are at least two states  of clouds with different densities.
Based on the detailed observational results, Bergeron et al. (1994) 
constructed a two-phase photo-ionized cloud model.
The inner core zone which exhibits Mg {\small II} absorption doublet has an   
 intermediate dimension ($\sim$ 7 kpc) with a density of 
$n_{\rm H} \sim $6$\times$10$^{-3}$ cm$^{-3}$  and a temperature 
of $T \sim$ 1.2$\times$10$^{4}$ K.
The outer O {\small VI} phase is homogeneous, with a lower density 
$n_{\rm H} \sim$ 3$\times$10$^{-4}$ cm$^{-3}$, 
with a high temperature of $T \sim$ 2.5$\times$10$^{4}$ K  
and of a large extent ($\sim$ 70 kpc).
The total mass of the gas is 10$^{9} M_{\solar}$ 
and the common metal abundance is about half the solar value. 
The average neutral hydrogen 
fraction is HI/H =3$\times$10$^{-3}$ at the core  and the neutral and ionized 
 hydrogen 
column density is $N_{\rm H}=$1.7$\times$10$^{20}$ cm$^{-2}$. 
Considering  the similarity in the size, temperature and high metal abundance,
a link between the O {\small VI} phase gas 
and a group of galaxies  is suggested 
(Bergeron et al. 1994).  
A Gunn-Peterson test using {\it HST FOS} detect an absorption trough 
with $\tau_{\rm GP-HI} =$ 0.12$\sim$0.14, which corresponds to an 
over-density by a factor of 3--7 in the intergalactic 
medium near $z \sim $0.8 
(Khersonsky et al. 1997).
 
If there is such a hot and dense intervening cloud in the line of sight, 
ionized heavy elements could produce  
absorption features in the  X-ray spectrum.
We observed PKS2145+067 with {\it ASCA} (Tanaka et al. 1994) 
to search for an  evidence of this hot intervening 
cloud using X-ray absorption features.
The observation  was performed on  5th December 
1998 with a net exposure time of 36 ksec with GIS and 29 ksec with  SIS,  
respectively. The observation mode was 
nominal PH mode for GIS (Ohashi et al. 1996, Makishima et al. 1996) 
 and 1CCD FAINT mode for SIS. 
In this paper, we assume 
 $H_{\rm 0}=$50 km sec$^{-1}$ Mpc$^{-1}$ and  $q_{0}=$0.5 
  and the solar number abundances of the elements are  followed by 
Anders \& Grevesse (1989).

\section{Analysis \& Results}
\subsection{Energy Spectrum  of PKS2145+067}
The source  has been clearly detected 
with  all the four detectors of {\it ASCA}. 
To make the energy spectra, 
we applied the standard data reduction criteria for  the data and 
integrated  the photons within a radius of 6 arcmin from the source.
Then 
we subtracted 
recent blank sky data  in the same observation mode 
to correct for both  the cosmic X-ray background and non X-ray background.
For the GIS background, 
a 80 ksec deep survey observation of the Lockman hole carried out 
in November 1998  (PI:Y. Ishisaki) is used. 
As for the SIS background, we used  1CCD FAINT mode data taken 
in the L1157 observation in September 1998 (PI:T. Furusho), in which  L1157 
dark cloud containing a class 0 proto-star 
has not been detected in X-ray band (Furusho et al. 2000). 
Since the background data were taken within 
3 months before the QSO observation, 
the long-term variation of the detector background and performances does not 
cause a problem  in the subtraction process.
In the L1157 data, selection of 
the same event discrimination level yields a  
net exposure time of the background to be  10.9 ksec.
However, the data  still have better  photon statistics than using 
the annular region  of the on-source SIS data 
because we can maximize the on-source integration area. 
In both GIS and SIS cases, S/N ratios are larger than 5 in all energy bands, 
and the total flux uncertainty due to the fluctuation of the background 
is less than 1\%.

The derived  spectra after the background subtraction are shown in figure 1. 
To avoid the uncertainty in 
the SIS response matrix which tends to cause a systematic excess absorption 
of a few times 10$^{20}$ cm$^{-2}$ (Cappi et al. 1998) due to the 
drop of the detection efficiency below 1 keV 
(see web pages of ASCA GOF, 
http://heasarc.gsfc.nasa.goc/docs/asca/watchout.html), 
we use only the data between 1.0 keV and 10.0 keV for SIS,
while the GIS data are used between 0.6 keV and 10.0 keV.
Spectral fitting package XSPEC ver 10.0 
is used for the model fitting throughout this paper.
We fit the  SIS and GIS data simultaneously and  minimize the 
sum of the $\chi^{2}$ assuming  an absorbed power-law model.

As for the absorber model, 
we first assume the interstellar matter in our Galaxy represented by 
  a XSPEC   ``wabs''  model at $z=0$, 
which is   the 
photoelectric absorption with  a metal abundance of one solar 
(Morrison \& McCammon 1983). 
All the data from SIS and GIS are consistent with each other and 
well described by a power-law with  interstellar absorption.
The resultant parameters are summarized in table 1.
The  absorption column density $N_{\rm H}$ is 
consistent with the Galactic value of 4.8$\times$10$^{20}$ cm$^{-2}$
(Dickey \&  Lockman 1990).
A 90 \% upper limit for 
 the equivalent width of  an iron-K emission line, assuming a narrow 
($\sigma <$ 500 eV ) Gaussian line at 
6.4 keV in the quasar  frame ($z=$0.990) is 50.4 eV.
The total flux between 2 and 10 keV in the observer frame is 
1.3$\times$10$^{-11}$ erg   cm$^{-2}$ sec$^{-1}$ which corresponds to 
 the luminosity 
 $L_{\rm X :2-10keV}=$ 2.5$\times$10$^{46}$ erg sec$^{-1}$ 
assuming $H_{\rm 0}=$50 km sec$^{-1}$ Mpc$^{-1}$ and $q_{0}=$0.5.  

\begin{table*}[t]
\begin{center}
Table~1.\hspace{4pt} The spectral parameters of PKS2145+067
\end{center}
\vspace{6pt}
\begin{center}
\begin{tabular}[t]{l|l|l|l|l}
\hline \hline
\multicolumn{5}{l}{wabs$\times$power-law} \\ \hline
\multicolumn{2}{l|}{$N_{\rm H} (z=0)$} & $\Gamma$ & $F_{\rm X:(2-10keV)}$ & $\chi^{2}$/dof \\ 
\multicolumn{2}{l|}{(10$^{20}$~cm$^{-2})$} &  & (erg cm$^{-2}$sec$^{-1}$) &  \\ \hline   
\multicolumn{2}{l|}{4.1 $^{+2.7}_{-2.6}$}& 1.63$\pm$0.04 & 1.3$\times$10$^{-11}$ & 792.7/853 \\ 
\hline \hline
\multicolumn{5}{l}{wabs$\times$zwabs($z=0.791$)$\times$  power-law} \\ \hline
$N_{\rm H}$($z=0$)  & $N_{\rm H}$($z=0.791$) & $\Gamma$ & $F_{\rm X:(2-10keV)}$ & $\chi^{2}$/dof \\ 
(10$^{20}$~cm$^{-2}$) & (10$^{20}$~cm$^{-2}$) &   & (erg~cm$^{-2}$sec$^{-1}$) &  \\ \hline   
4.8(fixed) & $<8.5$ & 1.64$^{+0.03}_{-0.02}$ & 1.3$\times$10$^{-11}$ & 792.6/853 \\ 
\hline \hline  
\end{tabular}
\end{center}
\end{table*}

\subsection{Upper limit for the absorption at z=0.791}
We added an additional absorption component at $z=$0.791
in the model and fit the spectra again. 
The assumed model is described by a formula; 
wabs $\times$ zwabs $\times$ a power-law.
The ``wabs `` model again represents  the absorption by the interstellar 
matter in our Galaxy and the column density $N_{\rm H}$ is fixed to the 
Galactic value of $N_{\rm H}=$ 4.8$\times$10$^{20}$ cm$^{-2}$. 
The ``zwabs'' model is a redshifted ``wabs'' model, which stands 
 for the absorption by neutral  matter with an abundance of 
1 solar.  
In this fit, we fixed the redshift of ``zwabs'' to be  $z=$0.791 
assuming its association to 
the Lyman limit clouds. 
The fitting results are also listed in table 1. 
The  90 \% upper limit for the absorption column at $z=0.791$ is 
8.5$\times$10$^{20}$ cm$^{-2}$.
A contour plot of the photon index $\Gamma$ vs. 
the absorption column density $N_{\rm H}$ at $z=0.791$ is shown in figure 2.

For the next step, we assumed 
that the metal abundance in the absorbing matter at 
$z=0.791$ is less than the solar level. 
We adopted a variable abundance 
photoelectric absorber 
using the XSPEC model of ``zvphabs'' (Balucinska-Church \& McCammon 1992), 
therefore, the model spectrum is described by the following formula; 
wabs($N_{\rm H}=$4.8$\times$10$^{20}$
cm$^{-2}$ fixed) $\times$ zvphabs ($z=0.791$) $\times$ power-law.
In this fit, all the elements are assumed to have a 
common abundance, except for He which has   an abundance of 1 solar.
In figure 3, we plot  upper limits of the absorption column density as 
a function of the metal abundance.
If the metal abundance is 0.5 solar, the 90 \% upper limit of the X-ray 
absorption column is 1.6$\times$10$^{21}$ cm$^{-2}$. 
This  does not contradict with the two-phase model 
($N_{\rm H}=1.7\times 10^{20}$ cm$^{-2}$) by 
Bergeron et al. (1994). 

We further  searched  for  absorption features due to discrete elements, 
for which {\it ASCA} data have good sensitivity.
We added  a redshifted absorption edge of Si.
In the photo-ionized cloud model of Bergeron et al. (1994), the most dominant 
population of Si is Si {\small VII} in the hot halo  and Si {\small III} 
in the core.
These ions produce 
a Si {\small III}  edge at 1852 eV (in the rest frame) of 
Si {\small III} and 
a Si {\small VII}  edge at  2001 eV which can be tested 
in the {\it ASCA} sensitivity band.
Again, we fitted the energy spectra with a absorbed power-law and an edge 
structure model given by the  following formula ;
wabs ($N_{\rm H} =$4.8$\times$10$^{20}$ cm$^{-2}$ fixed) 
$\times $ ``zedge'' (redshifted K-edge) $\times$ power-law .   
The 90 \% upper limits 
for  the optical depth of the edges are 
$\tau <$0.15 for 1852  eV (at $z=$0.791) Si {\small III} edge and 
$\tau <$0.08 for 2001 eV Si {\small VII}  edge.
These  optical depths correspond to column densities  of  
9.7$\times$10$^{17}$ cm$^{-2}$ and 6.3$\times$10$^{17}$ cm$^{-2}$ 
for Si {\small III} and Si {\small VII}  ions (Verner et al. 1994),
or assuming 0.5 solar abundances, the upper limits for the 
total hydrogen column density 
of $5.4\times 10^{22}$cm$^{-2}$ and $3.5 \times 10^{22}$cm$^{-2}$, 
respectively.

\subsection{Comparison with previous  X-ray observations}
 We compared the flux of  PKS2145+067 with the previous 
X-ray results. {\it Einstein IPC} observed the quasar 
in May 5th 1980. Wilkes et al. (1994) report
the flux between 0.16 and 3.5 keV to be 
3.8$\times$10$^{-12}$ erg~cm$^{-2}$ sec$^{-1}$ 
assuming a photon index of 1.5.
This  is about 1/3 of 
the level, $F_{\rm X:0.16-3.5keV}$ = 
1.1$\times$10$^{-11}$ erg~cm$^{-2}$ sec$^{-1}$,
estimated from the extrapolated  spectrum  
of the {\it ASCA} observation (wabs $\times$ power-law) 
without excess absorption.  
{\it ROSAT PSPC} also observed the quasar  on 9th   May  1991 and 
the flux between 0.1 and 2.0 keV is 
3.3$\times$10$^{-12}$ erg~cm$^{-2}$ sec$^{-1}$ (Perlman et al. 1998), 
which is about half of the extrapolated {\it ASCA} level
($F_{\rm X:0.1-2keV}$ = 
7.1$\times$10$^{-12}$ erg~cm$^{-2}$ sec$^{-1}$).

These flux comparison suggests two possibilities; the first is that 
the X-ray luminosity of 
PKS2145+067 has become higher   in 10 years by a factor of 2$\sim$3.
The radio flux of the source was monitored 
 at 318 and 430 MHz between 1980 and 1994 and showed 
little variation only by  7 \% rms 
(Salgado et al.1999).   
Sambruna (1997) compile {\it ROSAT } results and report that 
typical X-ray flux variation of flat-spectrum radio quasars (FSRQs) 
on  timescales of months/years
does not exceed  a factor of 2,  
characterized by  a  typical amplitude of the order of 
  10 -- 30 \% with   no accompanying spectral changes.
If the X-ray flux of PKS2145+067 has really varied by  a factor of 3, 
this  amplitude is the largest among the  reported FSRQ variations. 

The second possibility is that the energy spectrum is   strongly  
cut off   below the {\it ASCA} energy limit of 0.6 keV. 
Unfortunately, the spectral information from {\it ROSAT} and 
{\it Einstein} observations is  not available. We added an excess 
absorption in the fitting model to suppress  the extrapolated 
flux in the low energy band to  be consistent with the previous results.
In this  case, however, the required excess absorption 
becomes larger than 10$^{22}$ cm$^{-2}$ using the zvphabs model 
at $z=0.791$ with a metal abundance of 0.5 solar. 
This  is significantly larger than our  upper limit of 
1.6$\times$10$^{21}$ cm$^{-2}$. 
So it seems unlikely that  the flux difference  between 
{\it ASCA} and the previous soft X-ray observations is 
 only caused by  a strong absorption associated  with the Lyman limit clouds. 
It is, therefore, more plausible that the intrinsic luminosity of 
PKS2145+067 has  increased by a factor of 2$\sim$3 in these 10 years.

\section{Discussion \& Conclusion}
{\it ASCA} observation of PKS2145+067 shows that the 0.6--10 keV 
spectrum  is well described 
by  a power-law model ($\Gamma$=1.6)  absorbed by the 
Galactic interstellar absorption. 
An  upper limit for  the $z=0.791$ absorbing cloud 
is $N_{\rm H}<$1.6$\times$10$^{21}$ cm$^{-2}$ (90 \% U.L.).

These results are consistent with the two-phase ionized cloud model 
by Bergeron et al. (1996). 
However, to obtain a simple view of the absorbing cloud,
we calculated  simple one-phase photoionization 
models using CLOUDY ver 94.00 
(Ferland 1993, van Hoof et al. 2000).
We  studied  the requirement for 
the UV radiation field which satisfies  the upper limit of 
the total hydrogen column density.
We assume the absorber cloud to be a  plane-parallel slab of constant density, 
illuminated on both sides by an  ionizing radiation field.
The density of the cloud is varied between 
1.0$\times 10^{-3}$ and  1.0$\times 10^{-5}$ cm$^{-3}$ and 
the calculation is stopped when
the  neutral hydrogen column density reaches 
 $N_{\rm H}=3.2\times10^{17}$ cm$^{-2}$.
The metal abundance is  assumed to be 0.5 solar.
Considering the uncertainty of the UV radiation field at $z=0.791$, 
the intensity of the UV flux at 912A is assumed to be 
0.3, 1, 3 $\times$ 10$^{-22}$ erg cm$^{-2}$sec$^{-1}$sr$^{-1}$Hz$^{-1}$
(Okoshi \& Ikeuchi 1996), 
and the energy index $\alpha$ is assumed to be $-0.5$ and $-1.0$ .
In these condition, the ionizing parameter $U$ ($= n_{\gamma}/n_{\rm H}$) 
takes the values  between  $-2.72< \log(U) <0.28$.
There are 4 cases of strong radiation field, 
for  the 912 $\AA$ intensity 
of 1 and 3 $\times$ 10$^{-22}$ erg cm$^{-2}$sec$^{-1}$sr$^{-1}$Hz$^{-1}$ and 
 the index $\alpha$ of $-0.5$ and $-1$.
In these cases, the neutral hydrogen fraction becomes too low and 
the inferred total hydrogen column density $N_{\rm H} $ exceeds   
1.7$\times$10$^{20}$ cm$^{-2}$ for the fixed neutral hydrogen of 
$N_{\rm H}=3.2 \times10^{17}$cm$^{-2}$.
In weaker radiation field cases, 
the cloud size becomes smaller  than 1 Mpc only when 
the hydrogen density is larger than 4$\times$10$^{-4}$cm$^{-3}$, 
or $\log (U) <-1.9 $.
The electron temperature is $1.4 \times 10^{4} <T_{e}< 2.2\times 10^{4}$ K 
and the dominant spices of Oxygen are O {\small III} and O {\small IV}.
As the hydrogen density 
is larger than 4$\times$10$^{-4}$cm$^{-3}$ , which is 
close to the typical values   at the center of 
clusters and  groups of galaxies, it is possible that the cloud undergoes a 
gravitational collapse.
If collisional ionization is going on in the absorbing cloud, 
the fraction of highly ionized ions 
would become large and they could 
be detected as X-ray absorption structures.

Comparison with the previous {\it Einstein IPC} and {\it ROSAT PSPC} 
observations indicates that 
 the X-ray luminosity of  PKS2145+067 
has increased by  a factor of 2$\sim$3 between 1991 and 1998 and 
that this  discrepancy is  not due to an  excess absorption.
Sambruna  (1997) studied  time variability of 10 FSRQs 
in  timescales of months--years and 
found that the maximum flux  change is a factor of 2 drop in 0836+710 
during $\sim$ 8 months.     
The large flux change in PKS2145+067 shows  that the X-ray luminosity 
of FSRQs can vary by more than a factor of 2 
in a long timescale of the order of 10 years.

The present  {\it ASCA} observation have shown    that 
an improved  sensitivity for  the X-ray absorbing matter would 
enable us  to study optically 
observed Lyman limit or metal line systems
and that a wide-band spectroscopy with good  energy resolution 
will bring us a unique science.
We hope that new X-ray instruments with superior energy resolution 
on board  {\it XMM-Newton} and  {\it Chandra}
will be able to detect the ionizing hot intervening cloud directly 
by absorption lines and edges 
and reveal the physical conditions such as density, temperature, 
metal abundances of the inter-galactic medium.

\vspace{1pc}
The authors thank Dr. Y. Ishisaki and Ms. T. Furusho 
for their kind allowance to use their data for the 
background study.

\section*{References}
\re 
Anders  E., Grevesse N. 1989, Geochem. Cosmochim. Acta, 53, 197

\re
Bahcall J.N., Bergeron J., Boksenberg A., Hartig G.F., 
Jannuzi B.T., Kirhakos S., Sargent W.L.W., Savage B.D. et al.
\ 1993, ApJS  87, 1

\re
Balucinska-Church M.,  McCammon D. 1992, Ap.J 400, 699 

\re
Bergeron J., Petijean  P., Sargent W. L., Bahcall J.N., 
Boksenberg A.,  Hartig G.F., Jannuzi B.T., Kirhakos S. et al. 
\ 1994, Ap.J. 436, 33

\re
Bergeron  J.,Boiss\'{e} P.  1991, A\&A 243, 344

\re 
Cappi M., Matuoka M., Otani C., Leighly K.M. 1998, PASJ 50, 213 

\re
Dickey J.M., Lockman F.J. 1990, Ann. Rev. Ast. Astr. 28, 215

\re 
Ferland G.J. 1993, International Rept., Univ. Kentucky, Dept. Phys. and Astron.

\re
Furusho T., Yamasaki N.Y., Ohashi T., Saito, Y. Voges, W. 2000, 
will appear in PASJ, 52 

\re
Kellermann K.I., Vermbulen R.C., Zensus J.A., Cohen M.H. 1998, AJ\ 115, 1295

\re
Khersonsky V.K., Turnshek D.A., Strub S.M. 1997, ApJ\ 491, 29

\re
Makishima K., Tashiro M., Ebisawa K., Ezawa H., Fukazawa Y., 
Gunji S., Hirayama M., Idesawa E. et al. 1996, PASJ  48, 171 

\re 
Morrison R., McCammon D.  1983, ApJ.\ 270, 119

\re
Ohashi T., Ebisawa K., Fukazawa Y.,Hiyoshi K., Horii M., 
Ikebe Y., Ikeda H., Inoue H., et al. 1996, PASJ  48, 157
 
\re 
Okoshi K., Ikeuchi S. 1996, PASJ 48, 441 

\re 
Padovani P. 1992, MNRAS\ 257, 404

\re
Perlman  E.S., Paolo P., Giommi P., Sambruna R., 
Jones L., Tzioumis A., Reynolds J.  1998, AJ\ 115, 1253
  
\re 
Salgado J.F., Altschuler D.R., Ghosh T., Dennison B.K., Mitchel K.J., 
Payne H.E. 1999, ApJS\ 120, 77

\re
Sambruna R.M. 1997, ApJ\ 487, 536 

\re
Tanaka Y., Inoue H., Holt S.S.\ 1994, PASJ 46, L37

\re
van Hoof P.A.M., Martin P.G., Ferland G.J. 2000, astro-ph/0001196 

\re
Verner D.A., Ferland G.J., Korista K.T. 1996, ApJ 465, 487 

\re 
Visvanathan N., Wills B.J. 1998, AJ\ 116, 2119

\re
Wilkes B.J., Tanabaum, H., Worrall D.M., Avni Y.,   
Oey M.S., Flanagan J. \ 1994, ApJ.S. 92, 53

\clearpage
\centerline{Figure Captions}
\bigskip

\psfig{file=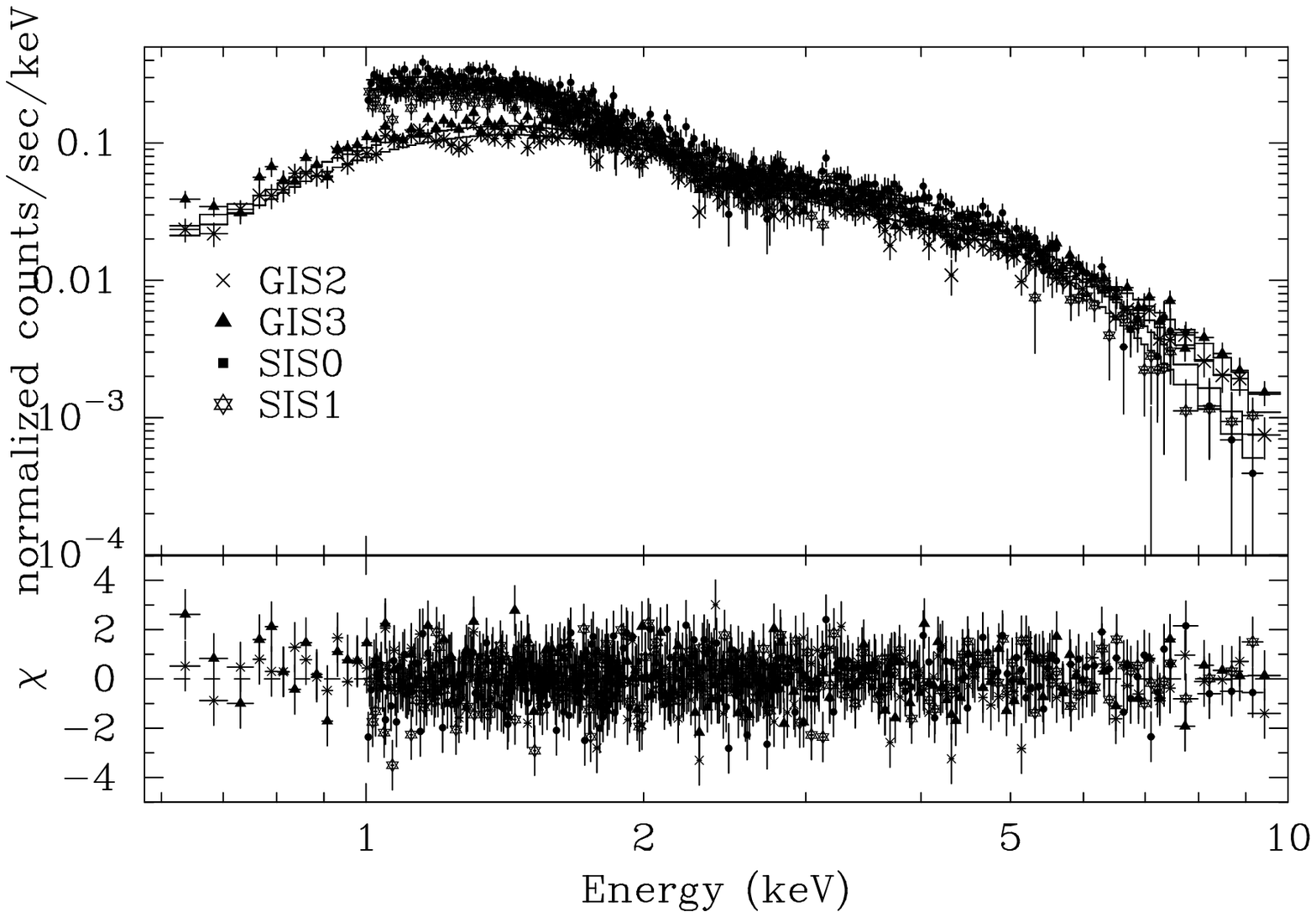,width=15cm}
\begin{fv}{1}
{}
{The pulse height  spectra of PKS2145+067 taken by SIS0, SIS1, GIS2, and GIS3, 
corrected for the background. The best-fit power-law model with interstellar 
absorption is shown.
The bottom panel shows the residuals of the fit.}
\end{fv}

\psfig{file=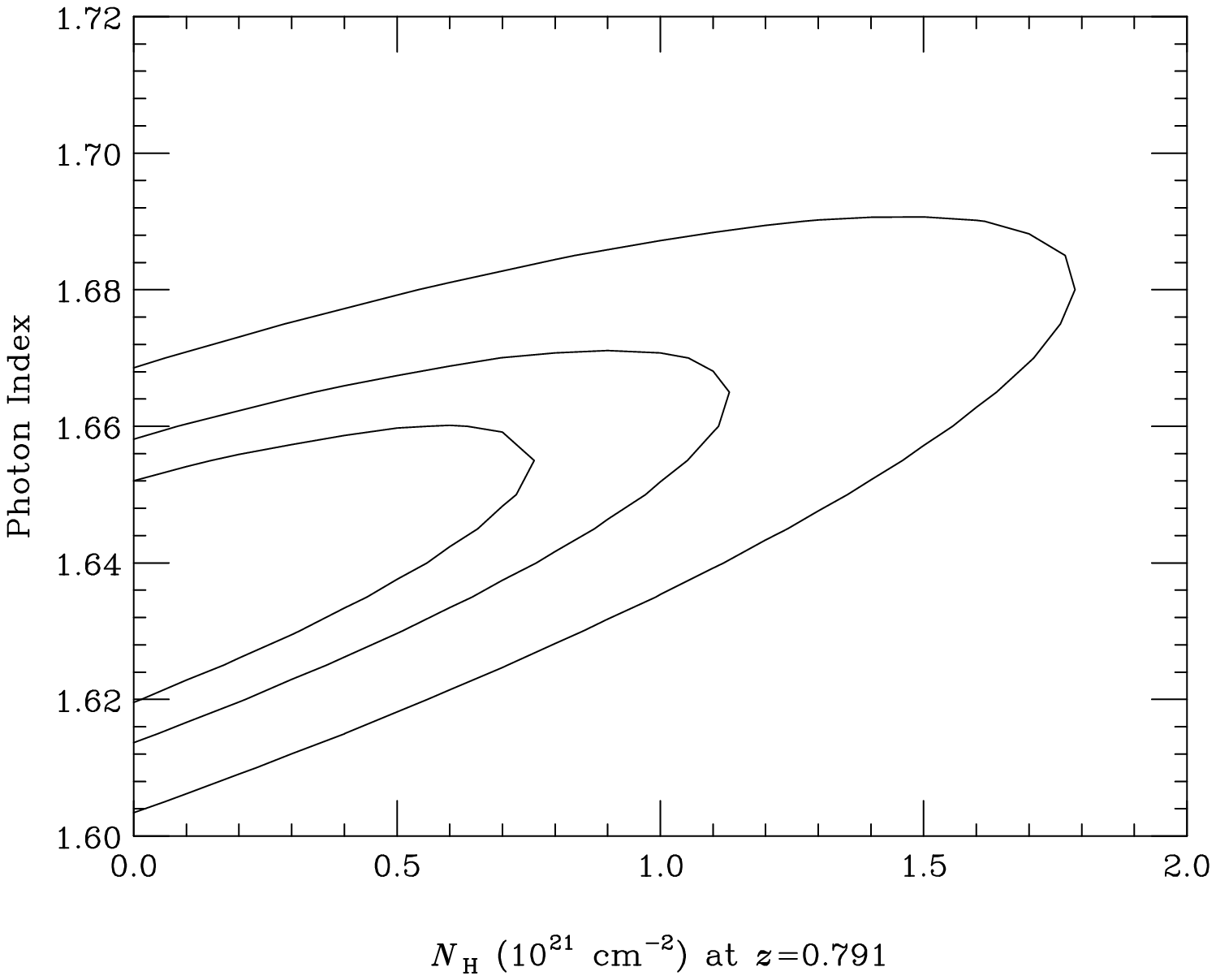,width=15cm}
\begin{fv}{2}
{}
{The Contour plot of the photon index $\Gamma$ vs. the absorption column 
density $N_{\rm H}$ at $z=$0.791. The absorbing gas is assumed 
to have a metal abundance of 1 solar. Lines show  67 \%, 90\%, and 99\% 
confidence levels.}
\end{fv}

\psfig{file=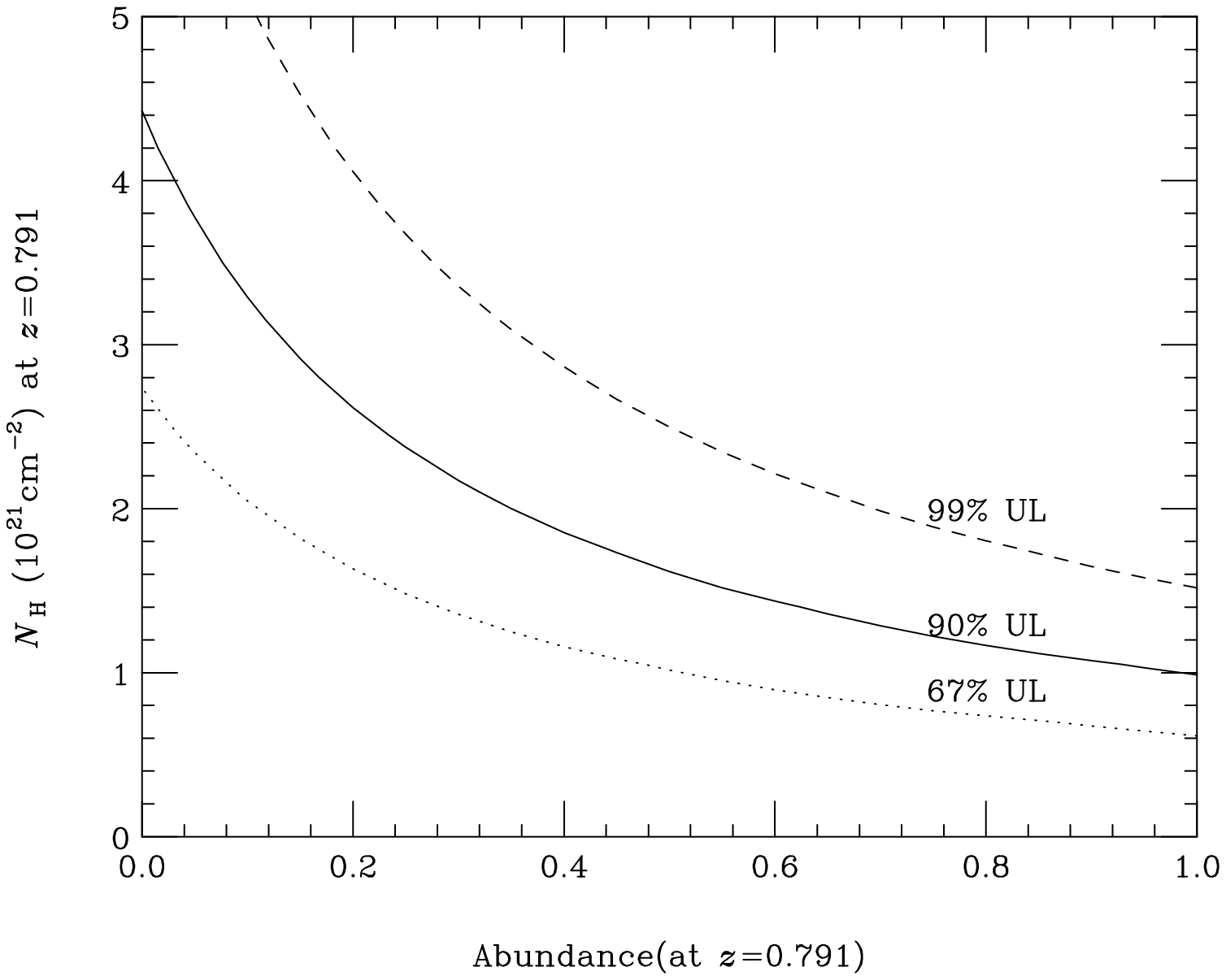,width=15cm}
\begin{fv}{3}
{}
{The upper limit of  the absorption column density at $z=$0.791 
for different metal abundances. Dashed, solid and dotted lines show 99\%, 
90\% and 67 \% confidence levels, respectively.}
\end{fv}

\end{document}